\documentclass[pra,showpacs,twocolumn]{revtex4}
\usepackage{times}
\begin{document}
\title{Canonical Quantum Teleportation}
\author{Sixia Yu}
\author{Chang-Pu Sun}
 \email{suncp@itp.ac.cn}
 \homepage{www.itp.ac.cn/~suncp}
\affiliation{The Institute of Theoretical Physics, Academia Sinica,
         Beijing 100080, P.R. China}
\date{April 26, 1999; Revised July 13, 1999}
\begin{abstract}
Canonically conjugated observables such as position-momentum and
phase-number are found to play a 3-fold role in the drama of the
quantum teleportation. Firstly, the common eigenstate of two
commuting canonical observables like phase-difference and
number-sum provides the quantum channel between two systems.
Secondly, a similar pair of canonical observables from another two
systems is measured in the Bell operator measurements. Finally,
two translations generated by the canonically conjugated
observables of a single system constitute the local unitary
operation to recover the unknown state. In addition, the necessary
and sufficient condition is presented for a reliable quantum
teleportation of finite-level systems.
\end{abstract}
\maketitle
The quantum teleportation \cite{bent}, a disembodied transmission
of quantum state, has been demonstrated in several experiments
both for finite-level systems \cite{exp} and continuous variables
\cite{vaid1,brn1,fur}. Along with the resulting discussions
\cite{brnd,vaid2} about its experimental realization, many other
aspects such as general schemes \cite{bent2,tor,stig} and some
applications \cite{chiral} of the quantum teleportation have also
been investigated. All these investigations so far emphasize
mainly on the {\em states} of the systems. In this Letter we shall
show the fundamental roles played by the canonically conjugated
($c.c.$) {\em observables} in the drama of the quantum
teleportation in order to reveal the physical contents of its
basic ingredients.

Generally speaking, the quantum teleportation consists of three
basic steps: (i) To prepare two systems in an
Einstein-Podolsky-Rosen (EPR) entangled state or a Bell state and
send them to two different places to establish a quantum channel;
(ii) At one place, to perform the so-called joint Bell operator
measurements with respect to one system involved in the EPR
entanglement and a third system at an unknown state to be
transferred; (iii) At the other place, to perform necessary
unitary operations to the other system involved in the EPR
entanglement according to the outcomes of Bell operator
measurements. By this means the unknown state is transferred from
one place to another.

In the case of continuous variables, three similar systems 1,2 and
3 are considered, which are described by canonical observables
$\hat x_a, \hat p_a$ $(a=1,2,3)$ satisfying canonical commutation
rules
\begin{equation}\label{ccm}
[\hat x_a,\hat p_b]=i\delta_{ab}, \quad (a,b=1,2,3).
\end{equation}
System 1 and 2 are prepared in a common eigenstate of the
position-difference $\hat x_1-\hat x_2$ and the momentum-sum $\hat
p_1+\hat p_2$ corresponding to eigenvalues $x_{12}$ and $p_{12}$
\cite{brn2}:
\begin{equation}\label{c1}
|x_{12};p_{12}\rangle= e^{-i\hat p_1\hat
x_2}|x_{12}\rangle_1\otimes|p_{12}\rangle_2,
\end{equation}
where $|x_{12}\rangle_1$ is an eigenstate of $\hat x_1$ with
eigenvalue $x_{12}$ and $|p_{12}\rangle_2$ is an eigenstate of
$\hat p_2$ with eigenvalue $p_{12}$. And system 3 is in an unknown
state $|\psi\rangle_3$ to be teleported to the first system.

Then a kind of Bell operator measurement measuring the
position-difference $\hat x_2-\hat x_3$ and the momentum-sum $\hat
p_2+\hat p_3$ is performed on systems 2 and 3. This measurement
projects systems 2 and 3 to one of the common eigenstates
$|x_{23};p_{23}\rangle$ of $\hat p_2+\hat p_3$ and $\hat x_2-\hat
x_3$ with $x_{23}$ and $p_{23}$ taking values on the real line
uniformly. Accordingly, system 1 is transformed into state ${\cal
O}_c ^\dagger|\psi\rangle_1$ where
\begin{equation}\label{c3}
{\cal O}_c=e^{-ip_{23}x_{12}} e^{-ip_{13}\hat x_1} e^{ix_{13}\hat
p_1}
\end{equation}
with $p_{13}=p_{12}-p_{23}$, $x_{13}=x_{12}+x_{23}$. At this
central stage, the measured observables are exactly two commuting
canonical observables: momentum-sum and position-difference.
According to the outcomes $x_{23}$, $p_{23}$ of the measurement
and values $x_{12}, p_{12}$ known from the state preparation, one
is able to perform unitary operation ${\cal O}_c$ to system 1. And
system 1 is then at the unknown state though no one knows what the
unknown state is.

Since $p_{12}$ and $p_{23}$ are the momentum-sums of corresponding
systems, $p_{13}=p_{12}-p_{23}$ is naturally the
momentum-difference between systems 1 and 3. Similarly
$x_{13}=x_{12}+x_{23}$ can be viewed as the position-difference
between systems 1 and 3. The unitary operation ${\cal O}_c$, being
made up of two successive translations up to a phase factor, has
therefore a natural physical meaning: it compensates the {\em
position-difference} and {\em momentum-difference} between systems
1 and 3. This obvious fact was already noticed in Ref.
\cite{vaid1} where the teleportation of continuous variables was
first proposed.

We see clearly that the $c.c.$ observables, position and momentum
in this case, play a 3-fold role in the drama of the quantum
teleportation of continuous variables. Firstly the common
eigenstate of two commuting canonical observables, e.g. the
position-difference and the momentum-sum, provides the quantum
channel between two systems. Secondly the same commuting canonical
pair of another two systems is measured in the Bell operator
measurement. Finally the $c.c.$ observables of a single system
generate two translations, which make up of the unitary operation
to recover the unknown state. So the quantum teleportation
deserves the name {\em canonical quantum teleportation}.

Given one pair of $c.c.$ observables one may design one possible
canonical quantum teleportation with exactly those three steps.
Notice that in the procedure of quantum teleportation the real
position and momentum cannot be used because localization of the
particle is required. In fact in the recent experimental
realization of the quantum teleportation of continuous variables
\cite{brn1} , a pair of $c.c.$ observables of the photon field,
phase quadrature and number quadrature, have been used.

At the very first look, in the case of the finite-level systems
those three steps of the quantum teleportation seem to be three
unrelated procedures: Bell states preparation, Bell operator
measurements \cite{brna} or nonlocal measurements \cite{vaid1} and
special unitary operations, whose physical meanings need
clarifying. We shall then demonstrate that there is also a pair of
$c.c.$ observables that plays the same 3-fold role for
finite-level systems. As it turns out, one observable is the
number operator and the other one is the phase operator of a
finite-level system.

For an infinite-level system as simple as a quantum harmonic
oscillator, a Hermitian phase operator does not exist
\cite{drc,lynch,index0}. After a series of efforts to solve this
problem \cite{newton,bant,luis0,pd} it was clear recently that the
quantum phase of a harmonic oscillator can only be described by
means of the phase-difference between two systems with a
rational-number-type of spectrum and the quantized
phase-difference obeys a quantum addition rule \cite{pd,index}.
Among the early approaches to this dilemma, the truncated Hilbert
space approach proposed by Pegg and Barnett \cite{bant}  describes
{\it in de facto} the phase variable of a finite-level system
instead of a harmonic oscillator with infinite many energy levels.
This approach was also investigated in some details by others
\cite{vds,luis1}.

For an $s$-level system $A$, the number operator ${\cal N}_A$ has
spectrum $Z_s=\{0,1,\ldots, s-1\}$ and its eigenstates
$|n\rangle_A$ with $n\in Z_s$ span the Hilbert space of the
system. In this Hilbert space, taking the phase window as
$[0,2\pi)$, one can define the exponential phase operator as
\begin{equation}
 e^{in{\cal P}_A}=\sum_{m\in Z_s}|m+n\rangle_A\langle m|,\quad n\in Z_s.
\end{equation}
Here the state $|ks+n\rangle_A$ is identified with the state
$|n\rangle_A$ whenever $k$ is an integer. This identification
seems to be trivial enough for a single system, but it is crucial
for the combination of number operators from different systems.
The so-defined exponential operator is obvious unitary which leads
to a Hermitian phase operator $\cal P_A$ with spectrum
$\Xi_s=\{2m\pi/s|m=0,1,\ldots,s-1\}$ and eigenstates
\begin{equation}
|\theta\rangle_A= \frac{1}{\sqrt s}\sum_{n\in Z_s
}e^{-in\theta}|n\rangle_A, \quad \theta\in\Xi_s.
\end{equation}

The motivation to define a Hermitian phase operator is, analogous
to the well-known canonical position and momentum, to find the
$c.c.$ partner for the number operator. However, the canonical
relationship between the quantum phase and number cannot be
explicitly manifested through their commutator. The quantum phase
and number have a very complicated commutator \cite{bant} due to
the fact that the phase variable has a curved configure space
because of its the periodicity, which is also the origin of the
rational-number-type of spectrum of quantized phase difference
\cite{pd}. Only when the unitary operations instead of Hermitian
observables are considered, dose the canonical relationship
between the phase and number  manifest itself \cite{manif}. As
shown explicitly in Eq.(\ref{c1}) and Eq.(\ref{c3}) it is also the
operations represented by unitary operators instead of the
observables represented by Hermitian operators that plays the main
roles in the case of continuous variables.

As is well known, the unitary operations generated by position and
momentum, which represent the translations in the momentum and
configuration spaces respectively, satisfy the Weyl form of
commutation relation
\begin{equation}\label{cc}
e^{ix\hat p}e^{ip\hat x}e^{-ix\hat p}e^{-ip\hat x}=e^{ixp}.
\end{equation}
This kind of relation indicates also the canonical relationship,
even more intrinsically than the commutator. This is because the
exponential phase and number operators satisfy also a similar
relation
\begin{equation}\label{dc}
 e^{i\theta{\cal N}_{A}}e^{in{\cal P}_{A}}
 e^{-i\theta{\cal N}_{A}}e^{-in{\cal P}_{A}}
 =e^{in\theta}.
\end{equation}
In this sense the quantum phase and number operator are $c.c.$
observables. The exponential phase-difference and
number-difference operators of two quantum harmonic oscillators
satisfy also this kind of relation which yield another pair of
$c.c.$ observables \cite{manif}.

As relation Eq.(\ref{cc}) indicates that the operator $e^{ix\hat
p}$ represents a translation by $x$ in the configuration space, so
the relation Eq.(\ref{dc}) ensures that the exponential phase
operator $e^{-in{\cal P}_{A}}$ represents also a translation by
$n$ (modular $s$) of the number. Similarly, the exponential number
operator $e^{i\theta{\cal N}_{A}}$ represents a translation by
$\theta$ (modular $2\pi$) of the quantum phase. These are exactly
the physical contents of these two unitary operations.

The quantum phase and phase differences were found to observe a
quantum addition rule \cite{pd}, which assures another quantum
phase or phase difference with the same kind of spectrum. The
quantum addition of phase operators ${\cal P}_A$ and ${\cal P}_B $
of two $s$-level systems $A$ and $B$, since they are commuting, is
simply $ {\cal P}_A\dot-{\cal P}_B\equiv{\cal P}_A-{\cal P}_B$
modular $2\pi$. Similarly, to preserve the spectrum of the number
operator, the quantum number-sum can be defined as $ {\cal
N}_A\dot+{\cal N}_B\equiv{\cal N}_A+{\cal N}_B$ modular $s$.
Because the quantum phase-difference and number-sum are commuting,
they possess common eigenstates
\begin{equation}\label{d1}
|\theta_{AB};n_{AB}\rangle= e^{-i{\cal N}_A{\cal P}_B}
|\theta_{AB}\rangle_A\otimes|n_{AB}\rangle_B,
\end{equation}
where $|\theta_{AB}\rangle_A$ is the eigenstate of ${\cal P}_A$
with eigenvalue $\theta_{AB}\in\Xi_s$ and $|n_{AB}\rangle_B$ is
the eigenstate of ${\cal N}_B$ with eigenvalue $n_{AB}\in Z _s$.
They form a complete and orthonormal basis of systems $A$ and $B$.
These two observables are measurable in the framework of nonlocal
measurements \cite{vaid1}.

Now that a complete analogue between the well-known $c.c.$
observables, position and momentum, and the less obviously {\it
c.c.} observables, quantum phase and number, has been established,
we can formulate the quantum teleportation of finite-level systems
in the same canonical manner. As a quantum channel of the quantum
teleportation of finite-level systems, systems $A$ and $B$ are
prepared in a common eigenstate $|\theta_{AB};n_{AB}\rangle$ of
their quantum phase-difference and number-sum.

Suppose that another $s$-level system $C$ is in an unknown state
$|\phi\rangle_C$ which will be teleported to the system $A$. To
this end we perform a joint measurement of the quantum
phase-difference ${\cal P}_{{B}}\dot- {\cal P}_{{C}}$ and the
number-sum ${\cal N}_{{B}}\dot+{\cal N}_{{C}}$  of the systems $B$
and $C$. With probability $1/s^2$, the total state of the whole
system
$|\Phi\rangle=|\theta_{AB};n_{AB}\rangle\otimes|\phi\rangle_C$ is
projected to state ${\cal O}_s^\dagger|\phi\rangle_A
\propto\langle \theta_{BC};n_{BC}|\Phi\rangle$ where
\begin{equation}\label{d3}
{\cal O}_s=e^{-in_{BC}\theta_{AB}} e^{-in_{AC}{\cal P}_A}
e^{i\theta_{AC}{\cal N}_A}
\end{equation}
with $\theta_{AC}=\theta_{AB}+\theta_{BC}$ and
$n_{AC}=n_{AB}-n_{BC}$ after the measurement. The number-sum
$n_{BC}$ takes value in $Z_s$ and the phase-difference
$\theta_{BC}$ takes values in $\Xi_s$ with equal probability,
which label the $s^2$ outcomes of the measurements.

After knowing these phase-differences $\theta_{AB}$, $\theta_{BC}$
and number-sums $n_{AB}$, $n_{BC}$, one can perform a unitary
transformation ${\cal O}_s$ to system $A$ so that the unknown
state of system $C$ appears at the other end of the quantum
channel. We note that operation ${\cal O}_s$ is made up of an
exponential phase operator and an exponential number operator up
to a phase factor. From the discussions above we know that these
two operations represent a phase translation by values
$\theta_{AC}$ and a number translation by values $n_{AC}$. Because
$\theta_{AC}$ can be regarded as the phase-difference and $n_{AC}$
as the number-difference between systems $A$ and $C$, the meaning
of these two unitary operations become now clear: before the
unknown state can be recovered the phase-difference and
number-difference between systems $A$ and $C$ must be compensated.

Consider the simple case of 2-level systems, where we identify
state $|0\rangle$ with $|\uparrow\rangle$ and state $|1\rangle$
with $|\downarrow\rangle$. As the quantum channel we prepare
systems $A$ and $B$ in the state as in Eq.(\ref{d1}) with
$\theta_{AB}=\pi,n_{AB}=1$. Four possible outcomes of the Bell
operator measurements on systems $B$ and $C$ are labeled by
phase-difference $\theta_{BC}=0,\pi$ and number-sum $n_{BC}=0,1$.
We can see that four corresponding unitary operations ${\cal O}_2$
in Eq.(\ref{d3}) applied to system $A$ are exactly the same as
those in Ref. \cite{bent}.

Canonical transformations, which preserve the canonical
commutators among observables as in Eq.(\ref{ccm}) or relations
such as Eq.(\ref{dc}) of corresponding unitary operations, can be
performed to $c.c.$ observables. Some canonical transformations
can result in some new forms of quantum teleportations. The
simplest case is to make a canonical transformation only to system
$B$, for example, ${\cal P}_B\to-{\cal P}_B$ and ${\cal
N}_B\to-{\cal N}_B$, which results a quantum teleportation as
follows. The quantum channel is a common eigenstate of ${\cal
P}_A\dot+{\cal P}_B$ and ${\cal N}_A\dot-{\cal N}_B$, e.g. state
\begin{equation}\label{is}
 |\Psi_{AB}\rangle=\frac1{\sqrt s}\sum_{m\in
 Z_s}|m\rangle_A\otimes|m\rangle_B
\end{equation}
corresponding to zero number-difference and zero phase-sum. The
observables measured in the second step are ${\cal P}_B\dot+{\cal
P}_C$ and ${\cal N}_B\dot-{\cal N}_C$. And the final operation
Eq.(\ref{d3}) to recover the unknown state remains unchanged. This
scheme is exactly the original teleportation of systems with more
than 2 levels discussed in Ref. \cite{bent}. One notes that when
$s=2$ the quantum phase-difference and number-sum are identical
with quantum phase-sum and number-difference respectively,
therefore these two teleportation schemes are identical in the
case of $s=2$.

Now we try to take a general pure state of systems $A$ and $B$ as
our quantum channel. Any normalized state can be expressed as
$T|\Psi_{AB}\rangle$ where operator $T$ acts only on system $A$
with ${\rm Tr}(T^\dagger T)=s$. Then we perform a general Bell
operator measurement on systems $B$ and $C$. This is equivalent to
projection to some orthonormal basis of systems $B$ and $C$
\begin{equation}
|k;l\rangle=\frac1{\sqrt s}\sum_{m\in
 Z_s}|m\rangle_B\otimes{\cal O}_{kl}|m\rangle_C,
\end{equation}
where $s^2$ operators ${\cal O}_{kl}$ act only on a single system
and satisfy the following normalization conditions
\begin{equation}\label{cond}
 {\rm Tr}\left({\cal O}_{kl}{\cal O}^\dagger_{k^\prime l^\prime}\right)
 =s\;\delta_{kk^\prime}\delta_{ll^\prime},
 \quad k,k^\prime,l,l^\prime\in Z_s.
\end{equation}
Numbers $k,l$ label all possible outcomes of the measurements.
Given outcomes $k,l$ of the measurements, appearing with equal
probability, system $A$ is found to be in state
\begin{equation}
s^2\langle k;l|T|\Psi_{AB}\rangle\otimes|\phi\rangle_C=T{\cal
O}_{kl}^\dagger|\phi\rangle_A,
\end{equation}
where operator ${\cal O}_{kl}$ is now acting on system $A$.  The
only requirement for a reliable quantum teleportation is therefore
to have $T{\cal O}_{kl}^\dagger$ unitary, which infers that $T$
must be reversible. From Eq.(\ref{cond}) one obtains
Tr$((T^\dagger T)^{-1})=s$, which is compatible with ${\rm
Tr}(T^\dagger T)=s$ iff $T$ is unitary. Therefore to have $T{\cal
O}_{kl}^\dagger$ unitary is equivalent to have all the operators
$T$ and ${\cal O}_{kl}$ unitary. This is the necessary and
sufficient condition for a reliable quantum teleportation. In
other words the quantum channel must be a maximum entangled state
and the measurements must be projections to maximum entangled
states. And the recovering operation at the final stage is simply
${\cal O}_{kl}T^\dagger$ depending on the outcomes of the
measurements.

As one wishes, from orthonormal bases $|k;l\rangle$ one can
construct two commuting canonical observables like
phase-difference and number-sum, whose common eigenstates are
exactly these bases. As a result, the measured observables in the
second step of the quantum teleportation may be different from the
observables determines the quantum channel. For example, the
quantum channel may be provided by the common eigenstate of the
quantum phase-sum and number-difference and the quantum
phase-difference and number-sum are the Bell operators. By this
means one can also teleport an unknown state from one place to
another. The general scheme discussed in Ref.\cite{stig} is
included here as a special example.

The continuous variables case can be analyzed similarly. Let us
fix our measurements at the second step to the projections to
states $|x_{23},p_{23}\rangle$. All the pure states that can be
used as quantum channel should have form
$\sum_{n=0}^{\infty}D^\dagger|n\rangle_1\otimes|n\rangle_2$ where
$D$ is an arbitrary unitary operator acting on system 1 only. The
operation at the final stage is $M^\dagger D$ where $M$ is a
unitary operator acting on system 1 with elements $\langle
m|M|n\rangle=\langle x_{23},p_{23}|m,n\rangle$ where
$|m,n\rangle=|m\rangle_1\otimes|n\rangle_2$ denote the number
state bases with $m,n$ going from zero to infinity. When the
elements of $D$ are taken as $\langle m|D|n\rangle=\langle
x_{12},p_{12}|m,n\rangle$, the teleportation of continuous
variables discussed at the beginning is regained. This discrete
formulation of the quantum channel upto a normalization constant
was noticed in Ref. \cite{enk}.

When one consider three quantum harmonic oscillators, although the
quantized phase-differences between each two of them are well
defined, it is impossible to perform a quantum teleportation using
the quantized phase difference and number-sum. This is because the
exponential phase operator of a single oscillator, which ought to
be employed to compensate a number-difference at the final stage
of the quantum teleportation, dose not exist.

In conclusion, the quantum teleportation is characterized by
$c.c.$ observables completely: The quantum channel is provided by
the common eigenstate of two commuting canonical observables, the
Bell operator measurement measures a similar pair of canonical
observables and the recovering operation consists of two
translations generated by the $c.c.$ observables. By applying
suitable canonical transformations to the $c.c.$ observables, one
can design new schemes of quantum teleportation. The necessary and
sufficient condition for a reliable quantum teleportation of
finite systems is to have a maximum entangled state as quantum
channel and the Bell operator measurements are projections to
maximum entangled states. The nonexistence of certain $c.c.$
observables makes the quantum teleportation using these variables
impossible. All these investigations concern the ideal quantum
teleportation. In the real experiments where non-ideal elements
must be considered, it becomes ambiguous how to characterize
quantum teleportation. In this aspect some efforts have been made
\cite{lam}. The attention to the roles played by the $c.c.$
observables in the drama quantum teleportation may help to
establish such kinds of criteria both for the continuous and
discrete variables.

One of the authors (S. Yu) gratefully acknowledges the support of
K. C. Wong Education Foundation for postdoctors, Hong Kong. This
work is also partially supported by the NSF of China.


\begin{thebibliography}{}
\bibitem{bent} C. Bennett, G. Brassard, C. Crepeau, R. Jozsa, A. Peres,
and W. K. Wooters, Phys. Rev. Lett. {\bf70}, 1895 (1993).
\bibitem{exp} D. Boschi, S. Branca, F. De Martini, L. Hardy, and S.Popescu, Phys. Rev. Lett. {\bf80}, 1121
(1998); D. Boumeester, J.-W. Pan, K. Mattle, M. Eibl, H.
Weinfurter, and A. Zeilinger, Nature (London) {\bf390}, 575
(1997); M. A. Nielsen, E. Knill and R. Laflamme, Nature {\bf 396},
5 November, 52 (1998).
\bibitem{vaid1} L. Vaidman, Phys. Rev. A {\bf 49}, 1473 (1994).
\bibitem{brn1} S. L. Braunstein and H. J. Kimble,
Phys. Rev. Lett. {\bf80}, 869 (1998).
\bibitem{fur} A. Furasawa, J. L. Sorensen,
S. L. Braunstein, C. A. Fuchs, H. J. Kimble and E. S. Polzik,
Science {\bf 282} 23 October, 706 (1998).
\bibitem{brnd} S. L. Braunstein and H. J. Kimble,
 Nature (London) {\bf394}, 840 (1997);
D. Bouwmeester, J.-W. Pan, M. Daniell, H. Weinfurter, M. Zukowski
and A. Zeilinger, Nature (London) {\bf394}, 841 (1997).
\bibitem{vaid2} L. Vaidman and N. Yoran, Phys. Rev. A {\bf59}, 116 (1999).
\bibitem{bent2}C. H. Bennett, G. Brassard, S. Popescu, B.
Schumacher, J. A. Smolin, and W. K. Wooters, Phys. Rev. Lett. {\bf
76}, 722 (1996).
\bibitem{tor} M. Tor, LANL e-print quant-ph/9608005.
\bibitem{stig} S. Stenholm and P. J. Bardroff,
Phys. Rev. A {\bf 58}, 4373 (1998).
\bibitem{chiral}C. S. Maierle, D. A. Lidar, and R. A. Harris,  Phys. Rev. Lett. {\bf 81}, 5928
(1998); M. S. Zubairy, Phys. Rev. A {\bf 58}, 4368 (1998).
\bibitem{brn2} G. J. Milburn and S. L. Braunstein, LANL e-print
quant-ph/9812018.
\bibitem{brna}S. L. Braunstein and A. Mann, Phys. Rev. A {\bf 51},
R1727 (1995).
\bibitem{drc}P. A. M. Dirac, Proc. R. Soc. London A {\bf114}, 193
(1927).
\bibitem{lynch} R. Lynch, Phys. Rep. {\bf367}, 256 (1995).
\bibitem{index0}K. Fujikawa, Phys. Rev. A {\bf52}, 3299 (1995).
\bibitem{newton}R. G. Newton, Ann. Phys. (N.Y.) {\bf124}, 327-346
(1980).
\bibitem{bant} D. T. Pegg and S. M. Barnett,
Phys. Rev. A {\bf39}, 1665 (1989).
\bibitem{luis0}A. Luis and L. L. S\`anchez-Soto, Phys.
Rev. A {\bf48}, 4702 (1993).
\bibitem{pd}S. Yu, Phys. Rev. Lett. {\bf 79}, 780 (1997).
\bibitem{index}S. Yu and Y. Zhang, J. Math. Phys. {\bf 39}, 5260 (1998).
\bibitem{vds}A. Vourdas, Phys. Rev. A {\bf41}, 1653 (1990).
\bibitem{luis1}L. Luis and L. L. S\`anchez-Soto,
 Phys. Rev. A {\bf46}, 1492 (1993).
\bibitem{manif}S. Yu, (in preparation).
\bibitem{enk}S. J. van Enk, LANL e-print quant-ph/9905081.
\bibitem{lam}T. C. Ralph and P. K. Lam, Phys. Rev. Lett. {\bf81},
5668 (1998).
\end{thebibliography}
\end{document}